\documentstyle[11pt]{article}
\begin{document}
\thispagestyle{empty}

\title{Non-rigid hole band in the extended t-J model}

\author{  R.\  Hayn \\
\em  Max Planck Arbeitsgruppe Elektronensysteme, \\
\em  Technische Universit\"at Dresden, \\
\em D-01062 Dresden, Federal Republic of Germany  \\
    \\  A.F.\ Barabanov \\
\em  Institute for High Pressure Physics, \\
\em  14202 Troitsk, Moscow Region, \\
\em  Russian Federation \\
   \\   J.\ Schulenburg  \\
\em  Institut f\"ur Theoretische Physik, \\
\em  Otto-von-Guericke-Universit\"at Magdeburg, \\
\em  D-39016 Magdeburg, Federal Republic of Germany \\
}

\maketitle
\thispagestyle{empty}

\newpage

\thispagestyle{empty}

\begin{abstract}
The dispersion of one hole in an extended $t$-$J$ model with additional
hopping terms to second and third nearest neighbours and a frustration term in
the exchange part has been investigated. Two methods, a Green's function
projection technique 
describing a magnetic polaron of minimal size and the exact diagonalization of
a $4*4$ lattice, have been applied, showing reasonable agreement among each
other. Using additional hopping integrals which are characteristic for the
CuO$_2$ plane in cuprates we find in the nonfrustrated case an isotropic
minimum of the dispersion at the point $(\pi/2,\pi/2)$ in $k$-space in good
coincidence with recent angle-resolved photoemission results for the
insulating compound Sr$_2$CuO$_2$Cl$_2$. Including frustration or finite
temperature which shall simulate the effect of doping, the dispersion is
drastically changed such that a flat region and an extended saddle point may
be observed between $(\pi/2,0)$ and $(\pi,0)$ in agreement with experimental
results for the optimally doped cuprates. 
\end{abstract}

\newpage

\pagestyle{plain}
\setcounter{page}{1}

\section{Introduction}

    It is well known that many properties of the high-$T_c$ superconductors 
(HTSC's) are determined by the spectrum of holes in the doped 
two-dimensional (2D) CuO$_2$ planes.
Recent experiments by angle resolved photoemission (ARPES) reported a 
small band width providing evidence for the existence of strong correlations
in cuprates [1-5]. Moreover they indicate the existence of a
flat dispersion 
region close to the Fermi surface for the optimally doped cuprates Bi2212, 
 Bi2201, Y123 and Y124 (for the notation see \cite{z5}). The flat region of
the quasiparticle band is located  
close to the point $(\pi, 0)$ in the irreducible part of the Brillouin zone
(where the lattice 
constant has to be set to unity $a=1$) and it has the form of an extended  
saddle point between $(\pi, 0)$ and $(\pi/2,0)$. The existence of
the flat band  
region leads to the appearence of a strong Van-Hove (VH) singularity which 
is close to the Fermi level. This circumstance is often used by different 
theories for the explanation of the high superconducting transition
temperature [3, 6-9]. Then, the existence of the
optimal value of doping 
is naturally connected with the movement of the chemical potential over the 
position of the VH singularity when the hole density is increased. So, the
investigation of the hole spectrum in the CuO$_2$ plane is of great importance 
for a microscopic theory of HTSC. 
\par
Important new results were obtained recently by ARPES experiments for 
the insulating, antiferromagnetic compound Sr$_2$CuO$_2$Cl$_2$ \cite{z9}. 
These results correspond to the situation of one hole in the  
quantum antiferromagnet and they serve as a test for different theoretical 
approaches. It was found \cite{z9} that the bottom of the hole band is close
to the  
point $(\pi/2,\pi/2)$ and the band width is of order 0.3 eV. It is 
essential,
that the effective mass was shown to be isotropic at the bottom of the band.
\par
One generally believes that the $t$-$J$ model 
describes qualitatively
the hole spectrum of the CuO$_2$ plane in HTSC. There are numerous studies of
the hole spectrum in the framework of the $t$-$J$ model based on the exact
diagonalization  
of small clusters \cite{1,2}, the self-consistent Born approximation
\cite{3,4} or a ``string'' ansatz for the hole wave function \cite{5}. One can
either start from the two-sublattice N\'{e}el-type state \cite{3,4,5}
or 
a spin rotational invariant spin liquid state \cite{6} and obtains
qualitatively the same result: the hole motion occurs mainly on one
sublattice, i.e.\ the dispersion relation is dominated by an effective hopping
to next nearest neighbours with the minimum of the dispersion at
$(\pi/2,\pi/2)$. One finds a flat region of the dispersion near to the point
$(\pi, 0)$. That was used for an attempt to interpret the experimental data of
the optimally 
doped compounds \cite{Dagotto}, where a rigid-band picture was assumed (see
also \cite{Eder}). However, the anisotropic minimum of the pure $t$-$J$ model
is in contrast to the experimental result in Sr$_2$CuO$_2$Cl$_2$ which shows an
isotropic band bottom and a large energy difference between
$(\pi/2,\pi/2)$ and $(\pi,0)$.
\par  
Recently it was shown \cite{z10}  that the qualitative agreement with the 
experimental results \cite{z9}  may be improved in the framework of the
$t_1$-$t_2$-$J$ model, which 
takes into 
account the next-nearest-neighbour hopping $t_2$ of the hole. The inclusion 
of a $t_2$-term substantially improves the correspondence of the calculated 
spectrum near the band bottom with the experimentally observed isotropic 
minimum of $\varepsilon(k)$ at the point $(\pi/2,\pi/2)$. On the other hand,
the hole spectrum of the  
$t_1$-$t_2$-$J$ model does not reproduce the flat band region near the point
$(\pi,0)$. So, we have the following theoretical problem: the one-hole
approach in the pure $t$-$J$ model can reproduce the flat band region of the
optimally doped cuprates [1-5] but not the isotropic band
bottom of the insulating compound \cite{z9}. And vice versa, one hole in the
$t_1$-$t_2$-$J$ model leads to an isotropic band bottom but not to a flat band
region near $(\pi,0)$. One possible solution could be that we have to choose
different microscopic models for the different compounds. Here, we will
investigate the other possibility, whether it is possible that the hole band
is changed by doping so that a flat band region between $(\pi,0)$ and
$(\pi/2,0)$ arises.
\par
To simulate the effect of doping in a study of the one-hole motion there 
were recently
compared two possibilities, namely the influence of frustration and 
temperature \cite{z11}. The inclusion of frustration leads to a $t$-$J_1$-$J_2$
model, where $J_1$ and $J_2$ denote the antiferromagnetic exchange interaction
between  
nearest and next nearest neighbours. The influence of frustration was
investigated by two different methods, namely by a variational ansatz where the
spin-spin 
correlations in the frustrated Heisenberg model had been calculated by a spin
rotational invariant procedure \cite{12} and by the exact diagonalization of a 
$4*4$ 
lattice. 
It was found that the frustration shifts the minimum of the hole dispersion
from $(\pi/2,\pi/2)$ to the point $(\pi,0)$. Both
methods had a quite reasonable agreement where the variational method showed
the effect in a more pronounced way. Flat
dispersion regions were found in the nonfrustrated and frustrated cases. Without
frustration, the flat region occurs around $(\pi/2,\pi/2)$, whereas for
frustration $J_2/J_1 \stackrel{>}{\sim} 0.4$ it occures between $(\pi,0)$ 
and
$(\pi,\pi/2)$. Such a flat region is similar to the experiment
[1-5], but there it 
appears between $(\pi,0)$ and $(\pi/2,0)$. 
\par
The mentioned above investigations demonstrate that the extension of 
the $t$-$J$ model by the inclusion of additional hopping terms, frustration
or by taking  
into account a finite temperature may strongly modify the initial features of
the hole spectrum. But at the same time, none of these extensions alone can
lead to an adequate description of the experimental picture.      
\par
In the present paper we will investigate the hole spectrum in the framework of
the $t_1$-$t_2$-$t_3$-$J_1$-$J_2$ model, e.g.\ we will take  into account
the hole hoppings between first ($t_1$), second ($t_2$) and third ($t_3$) 
nearest neighbours and the frustration in the spin subsystem simultaneously.  
The additional hoppings $t_2$ and $t_3$ naturally appear in the Hamiltonian if
one reduces the well known three band Hubbard model of
the CuO$_2$ plane 
to an extended $t$-$J$ model \cite{JF,z13,z14}. Throughout the present paper
we 
will fix $t_2$ and $t_3$ to such values which can be derived from the
characteristic parameters of the three band Hubbard model for the CuO$_2$
plane which were given by Hybertsen {\em et al.\ } \cite{z12}. 
The frustration will be understood to simulate the doping as
proposed, for   
instance, in Ref.\ \cite{10}. The effect of a finite  temperature will also be 
discussed. 
\par
Two methods will be used to investigate the problem. The projection 
Mori-Zwanzig method \cite{z15} for two-time retarded Green's functions
will be employed to treat the hole excitation 
as a magnetic polaron of minimal size. Simultaneously we will use 
the exact diagonalization of
a $4*4$ lattice.  Within the projection method the dispersion is 
determined by static spin-spin 
correlation functions. These are calculated here in a spin rotational
invariant method \cite{12} (see also the zero temperature
version \cite{12a}). We describe the state of the magnetic 
subsystem as a spin liquid state in contrast to the widely used two-sublattice
N\'{e}el-type state due to several reasons. At first, our choice gives the
possibility to avoid the degeneracy of the hole-spectrum between the points
$(0,0)$ and $(\pi,\pi)$ which always occurs in the N\'{e}el-type
state. Second, the spin liquid state does not contradict 
the Mermin-Wagner theorem \cite{13} which forbids magnetic order in two
dimensions 
for any finite temperature. So it gives the possibility to investigate the
one-hole motion at non-zero temperatures. 
Of course, the Mori-Zwanzig projection method is an approximate one. 
Therefore, it is reasonable to investigate the problem by the exact
diagonalization method and to compare the results of two independent
procedures. 
\par
The paper will be organized as follows. At first, in Sec.\ 2 we will present
the spin polaron approach in the framework of the projection method. 
Then we show the results of both methods for zero temperature and
increasing frustration (Sec.\ 3) and discuss the influence of
temperature. Finally we present our conclusions.

\section{Small spin-polaron approach for the hole spectrum}
We consider an extended $t$-$J$ model on a square lattice with hoppings
between first, second and third nearest neighbours and with a frustration
term in the exchange interaction. The Hamiltonian of the model is given by
\begin{equation}
H = H_t + H_J =
- \sum_{i \sigma l} t_{l} \ X_i^{\sigma 0} X_{i+l}^{0 \sigma}
+ \frac{1}{2} \sum_{i m} J_{m} \ \vec{S}_i \cdot \vec{S}_{i+m}
\; ,
\label{1}
\end{equation}
where
\begin{equation}
t_{l}= \left\{ \begin{array}{ll}
t_1>0    &   \mbox{if $l$ is a nearest neighbour vector} \\
t_2      &   \mbox{if $l$ is a second nearest neighbour vector} \\
t_3      &   \mbox{if $l$ is a third nearest neighbour vector} \\
0        &   \mbox{else}
\end{array}
\right.
\label{1a}
\end{equation}
and
\begin{equation}
J_{m}= \left\{ \begin{array}{ll}
J_1>0    &   \mbox{if $m$ is a nearest neighbour vector} \\
J_2>0    &   \mbox{if $m$ is a next nearest neighbour vector} \\
0        &   \mbox{else}
\end{array}
\right. \; .
\label{1b}
\end{equation}
The Hamiltonian is expressed in terms of the Hubbard projection operators
which exclude the double occupancy at site $i$: acting on the vacuum state
$X_i^{\sigma 0}$ creates the electron (annihilates the hole) at site $i$ with
spin $S=1/2$ and spin projection $\sigma/2$ ($\sigma=\pm 1$ is a spin index). 
The following operator equations are fulfilled
\begin{equation}
\begin{array}{lll}
X_i^{\sigma 0}=c_{i \sigma}^{\dagger} ( 1 - n_{i -\sigma}) \; , &
X_i^{\sigma \sigma}=n_{i \sigma} ( 1 - n_{i -\sigma}) \; , &
X_i^{\sigma -\sigma}=c_{i \sigma}^{\dagger} c_{i -\sigma}\; , \\
n_{i \sigma}=c_{i \sigma}^{\dagger} c_{i \sigma} \; , &
S^\sigma_i = X^{\sigma -\sigma}_i \; , &
S^z_i = \frac{1}{2} \sum_{\sigma}
\sigma X^{\sigma\sigma}_i
\; , \\
X_i^{00} + \sum_{\sigma} X_i^{\sigma \sigma}=1 \; , &
X_i^{\lambda_1 \lambda_2} X_i^{\lambda_3 \lambda_4} = 
X_i^{\lambda_1 \lambda_4} \delta_{\lambda_2 \lambda_3} \; ,  & \\
\end{array}
\label{2}
\end{equation}
where 
$\lambda_n = 0$ or $\sigma$ in the last equation.
The hopping parameters $t_l$ in (\ref{1}) are not free but have to be
determined by a reduction procedure from the three band 
Hubbard model of the CuO$_2$ plane. We use here an analytical reduction
procedure, the so-called ``cell perturbation method'' \cite{JF,z13,z14}. For
the  
parameters of the three band Hubbard model we choose the values which
were found by Hybertsen {\em et al.\ } \cite{z12} for La$_2$CuO$_4$ by a
constrained density-functional calculation, namely:
$t_{pd}=1.3 \ eV, t_{pp}=0.65 \ eV,
\varepsilon_p-\varepsilon_d=3.6 \ eV, U_d=10.5 \ eV, U_p=4 \ eV$ and $U_{pd}=1.2
\ eV$ \cite{z12}. Band structure calculations for Sr$_2$CuO$_2$Cl$_2$
\cite{NFJ} show 
only few differences to the La-compound. Therefore, we expect that the
parameter values of Hybertsen {\em et al.\ } are also responsible for the
Sr-compound. From the reduction procedure (details will be published in Ref.\
\cite{z14}) we found the following 
hopping parameters: $t_1=498 \ meV, t_2=-41 \ meV$ and $t_3=77 \ meV$. They are
within the limits which have been found in \cite{z13} and also near to the
results of a numerical mapping in \cite{z12}. But in difference to Hybertsen
{\em et al.\ } we found it necessary to include the $t_3$-term also. In the
following we will choose $t_1$ as the unit of energy, i.e. $t_2=-0.08,
t_3=0.15$. So you see that the additional hopping terms are rather small, but
nevertheless essential as will be shown below. The reduction procedure gives
also the possibility to calculate the exchange energy $J_1$. That is however
less important for our purpose, since we will show that the results for
different values of $J_1$ can be roughly transformed into each other by
scaling the bandwidth with $J_1$. 
\par
Let us discuss the one-hole excitations on the background of the half-filled
rotationally invariant singlet states $\left| \Psi \right>$ of the pure spin
system $H_J$. We are interested in the energetically lowest
branch of these excitations and it is known that the bottom of its band is
represented by the states with total spin $S_{tot}=1/2$. 
We will treat the problem within the framework of the spin polaron of
small radius. That means that we restrict ourselves to spin excitations in the
immediate neighbourhood of the hole. 
It may be seen
that the full basis of such excitations is given by the following 5 basis
operators ${\phi_i^a}^{\dagger}$
with
$a=0,1,\dots 4$:  
\begin{equation}
\phi_i^{0}=X_i^{\sigma 0} \quad , \quad
\phi_i^{a}=\sum_s X_{i-a}^{\sigma s} X_i^{s 0} \quad (a=1,\dots,4)
\; ,
\label{3}
\end{equation}
where we will use the following notation hereafter: the small latin letters
$a$ and $b$ 
denote either a number between $0$ and $4$ or the corresponding lattice
vector
\begin{equation}
\begin{array}{ccc}
0 \leftrightarrow (0,0) & 1 \leftrightarrow (1,0) &
3 \leftrightarrow (-1,0) \\
 & 2 \leftrightarrow (0,1) &  4 \leftrightarrow (0,-1) \\
\end{array}
\label{4}
\end{equation}
in a synonymous way. Any vector ${\phi_i^a}^{\dagger} \left| \Psi \right>$
corresponds to a one hole state with $S_{tot}=1/2$ and $S_{tot}^z=-\sigma/2$, 
where
$S^{tot}$ and $S_{tot}^z$ are the spin and its projection of the total
system. 
\par
To obtain the one-hole spectrum $\varepsilon(k)$ we use the two-time retarted
matrix Green's function $G^{a b} (t,k)$ for the Fourier transformation of the
operators $\phi_i^a$:
\begin{equation}
G^{a b} (t,k) = \langle\!\langle 
\phi_k^a (t);
{\phi_k^b}^{\dagger} \rangle\!\rangle  =
-i \Theta(t) \langle \{ 
\phi_k^a (t);
{\phi_k^b}^{\dagger} (0) \} \rangle \; ,
        \label{5}
\end{equation}
\begin{equation}
\phi_k^a 
=\frac{1}{\sqrt{N}} \sum_j
\mbox{e}^{i k j} \phi_j^a \; ,
\label{6}
\end{equation}
where $\{ \ldots , \ldots \}$ denotes the anticommutator and where we have
used 
Zubarev's notation ~\cite{17}. 
\par
The equation of motion for the Green's function (\ref{5}) after going over
from time $t$ to frequency $\omega$ by Fourier transformation
$\langle\!\langle \phi_k^a ; {\phi_k^b}^{\dagger} \rangle\!\rangle_{\omega} $
has the form
\begin{eqnarray}
\omega 
\langle\!\langle \phi_k^a ; {\phi_k^b}^{\dagger} \rangle\!\rangle_{\omega}
&=& 
S^{a b}(k) +
\langle\!\langle \varphi_k^a ; {\phi_k^b}^{\dagger} \rangle\!\rangle_{\omega}
\; , \label{7a} \\
S^{a b}(k) 
&=& 
\langle \{ \phi_k^a, \phi_k^{b \dagger} \} \rangle \; ,
\quad
\varphi_k^a =
[ \phi_k^a, H ] \; ,
\label{7}
\end{eqnarray}
with the comutator $[ \ldots , \ldots ]$. 
To restrict ourselves to the above chosen set of 5 operators $\{ \phi_k^a \} $
(\ref{3}) 
we use the standard Mori-Zwanzig projection method \cite{z15} for the
operators $\varphi_k^a$:
\begin{eqnarray}
\varphi_k^a &\simeq & \sum_b \Omega^{a b} (k) \phi_k^b 
\; ; 
\quad
\underline{\Omega}(k)
=\underline{H}(k) \underline{S}^{-1}(k) \; ; \\ 
H^{a b} (k) &=& 
\langle \{ [ \phi_k^a , H ] , {\phi_k^b}^{\dagger} \} \rangle =
\left< \Psi \right| \phi_k^a  H   {\phi_k^b}^{\dagger} \left| \Psi \right> 
- E_0 S^{a b}(k) \; , 
\label{8}
\end{eqnarray}
where $E_0$ denotes the ground state energy of the spin system $H_J$ without
any hole. Note that the expression with the anticommutator in (\ref{8}) can be
simplified considerably for the one-hole problem. Using 
(\ref{8}) the eqn.\ (\ref{7a}) takes the following matrix form
\begin{equation}
\left( \omega \underline{E}
-\underline{H}(k) \underline{S}^{-1}(k) \right) \underline{G}
= \underline{S}(k) \; ,
\label{9}
\end{equation}
where $\underline{E}$ is a unit matrix. The quasiparticle spectrum
$\varepsilon(k)$ is determined by the poles of the 
Green's function $\underline{G}$ from the condition 
\begin{equation}
\det \left|
\varepsilon(k) \underline{S}(k) - \underline{H}(k) \right|=0 \; .
\label{10}
\end{equation}
The matrix 
elements of $\underline{H}$ and $\underline{S}$ may be calculated in a
straightforward way and the explicit expressions are given in the
Appendix. These matrix elements 
are expressed
through spin-correlation 
functions of the spin system for an undoped lattice which is described by the
frustrated $S=1/2$ Heisenberg model. As mentioned in the Introduction we treat
this model in the framework of the spherical symmetric approach which
gives values of pair spin-spin correlation functions at any value of
frustration parameter $J_2/J_1$ and temperature. 
Let us note that
the Ritz variational principle with the 5 basis vectors ${\phi_i^a}^{\dagger}
\left| \Psi \right>$  leads exactly to the same condition (\ref{10}) and the
same matrices $\underline{H}$ and $\underline{S}$ \cite{6,z11}. So, at least
for the case 
of the one-hole motion, the Mori-Zwanzig projection technique is identical to
the Ritz variational principle. 
\section{Numerical results and discussion}
As mentioned above we calculate the spectrum $\varepsilon(k)$ using the
hopping parameters $t_1$, $t_2$, $t_3$ which were obtained for La-cuprates
and take $t_1$ as the unit of energy. So we have the values $t_2=-0.08$ and
$t_3=0.15$. In the calculations of the magnetic polaron of minimal size we
choose $J_1=1$ since our relevant set of operators is too limited to describe
the small $J_1$ case. But we will see that one can reach the physically
relevant region by a simple rescaling. 
\par
Simultaneously with the Green's function method the
Lanczos exact diagonalization scheme is used as a
complementary method to calculate the
low-lying eigenstates for a square lattice of $4*4$ sites (with periodic
boundary conditions). These low-lying states are classified by its momentum
$k$ and the total spin. Here we will concentrate on the band with total spin
$1/2$. There are only few exceptions 
where the lowest states have spin $3/2$ and that happens only for higher lying
levels with momentum
$(\pi,\pi)$ or $(0,0)$. 
The Mori-Zwanzig projection method
gives the energy difference
between the lowest state with one hole for total spin $1/2$ and fixed momentum
and the ground state without any hole. 
The same energy difference is calculated for the $4*4$ lattice.
\par  
To clarify the importance of the additional hopping terms and frustration, in
Fig.\ 1 we represent the results for 
$J_1= t_1=1$ without frustration (Fig.\ 1a) and with $J_2=0.4$ (Fig.\ 1b). For
convenience we have chosen a finite temperature $T=0.2$ to calculate the
static spin-spin correlation functions \cite{12,12a}, 
but we have checked that the differences to the zero
temperature case can be neglected. 
One observes a quite
reasonable agreement between the Green's function method  and the exact
diagonalization data for these parameter values. As it is seen in 
Fig.\ 1a, in the absence of frustration, the spectrum 
demonstrates an isotropic band bottom close to the point $(\pi/2,\pi/2)$  in
accordance to the experimental dispersion of one hole \cite{z9}.   
The same result was obtained in Ref.\ 
\cite{z10} for the $t_1$-$t_2$-$J$ model where the following parameter values
were taken: $t_2= -0.35t_1$, $J= 0.3t_1$. 
Let us mention
that the isotropization of the band bottom in comparison with the $t$-$J$
model  may be obtained
in a rather wide range of parameters $t_2$ and $t_3$. For example, also the
values $t_2=-0.15t_1$ and $t_3=0.1t_1$ give a dispersion which is similar to
that one shown in Fig.\ 1a. 
It is important that the spectrum in Fig.\ 1a does not 
demonstrate a flat band region near the point $(\pi,0)$.  
\par
Before considering the spectrum of Fig.\ 1b let us remind that we
 suppose that the frustration simulates doping \cite{10}. Of course there is 
no full equivalence between doping and frustration. For example, the doped   
 $t$-$J$ model and the frustrated $J_1$-$J_2$ model give 
different results for the dynamical spin-spin structure factor and for the
spectrum of Raman scattering \cite{BGN}. 
Nevertheless, it is well known that both doping and 
frustration lead to a decrease of the magnetic correlation length.
 Furthermore, numerical calculations on finite lattices  
indicate the equivalence of the mentioned  models if we are interested in
  the static spin-spin correlation functions \cite{M}. Note that this is
especially relevant in our present Green's function method where the spectrum
$\varepsilon(k)$ is determined by the static spin-spin correlation functions
of the spin subsytem. 
\par
Fig.\ 1b represents the spectrum for the same energetical parameters as 
Fig.\ 1a, except the inclusion of a frustration term $J_2 = 0.4J_1$. The 
comparison of Figs.\ 1a and b indicates that the frustration dramatically 
changes the spectrum in the vicinity of the point $(\pi,0)$. It 
leads 
to the appearence of a flat band region close to that point. Moreover, this
flat  
band region has the form of an extended saddle point which stretches in the 
direction $(\pi/2,0)$-$(\pi,0)$. Such a band structure corresponds to the ARPES
results for  
optimally doped cuprates [1-5]. 
\par
So, we can describe the experimental results for insulating (no frustration,
Fig.\ 1a) and optimally 
doped compounds (Fig.\ 1b, the case with the frustration) by the same set of
hopping parameters. It is important that at the same time our approach  
demonstrates the non-rigid band behavior in the framework of a simple and
natural 
mechanism: the doping leads to the frustration in the spin subsystem and to
the variation of spin-spin correlation functions, and correspondingly
the alteration of these functions results in the non-rigid
behaviour of the spectrum. Or, in other words: the doping changes the state of
the magnetic background which strongly determines the form of the spectrum.
Let us mind that this mechanism doesn`t take into account the direct 
interaction between holes which will be of course important in the regime of
heavy doping.  
\par
In Figs.\ 2a and b we compare the exact diagonalization and the Green's
function projection method for 
several momenta in
dependence on the frustration $J_2/J_1$.   
The non-rigid band behaviour is clearly seen. Both figures demonstrate the
following common features. 
First, without frustration, the order of the levels coincides. Then,
the change of the energy due to frustration is quite similar for the 
lowest four levels.
The energy difference between the levels $(\pi/2,\pi/2)$ and $(\pi,0)$
decreases as the frustration increases, i.e. the level $(\pi,0)$ goes
 down with the respect to the band bottom (in the case $t_2 = t_3 = 0$ 
there is even an inversion of the levels  
and the band bottom occures close to $(\pi,0)$ \cite{z11}). One finds a
crossing for the 
levels   $(\pi/2,0)$ and 
$(\pi,\pi/2)$ at $J_2/J_1 = 0.17$ and $0.27$ for 
the projection and exact diagonalization methods, respectively. In addition,
the energy difference $\varepsilon(\pi,0)-\varepsilon(\pi/2,0)$ 
decreases in both methods, and both levels cross in the Green's 
function projection technique. That corresponds to
the appearance of the flat band region and the 
extended saddle point in the vicinity of the $(\pi,0)$-point. Of course, there
are also remarkable deviations of the Green's function method, especially for
the level $(\pi,\pi)$. But fortunately, that concerns only a high lying state.
\par
As a general tendency, the 
effect of frustration is more pronounced in the projection method and it seems
to be overestimated. 
This is already known for the static spin-spin correlation functions 
treated in the spherical symmetric approach \cite{12,12a} which is
incorporated in the present
calculation. This approach leads to a critical value $J_2/J_1=0.11$ for the  
second order transition between long range order 
and the spin-liquid state of the frustrated Heisenberg model \cite{12a} which
is rather low in comparison with other theories \cite{SZ,F3}.       
\par
The reasonable coincidence of the Green's function method with the exact
diagonalization data at $J_1=1$ is understandable
since  
for such large values of
$J_1$ the magnetic energy
stabilizes the size of the magnetic polaron and our concept of a polaron of
small radius is justified. Let us note that for $J_1\ll t_1$ the bandwidth
tends to be proportional to $J_1$ [11-15]. In principle this result
may 
also be obtained within our projection  procedure if we enlarge the polaron
size by enlarging the space of operators . To clarify the problem, 
in Fig.\ 2c we represent the dependence of the levels on the frustration 
for the same parameters  as in Fig.\ 2b, but for a more
realistic value of the exchange interaction $J_1=0.4$. Comparing Figs.\ 2b and
2c we observe that the bandwidth for $J_1=0.4$ is scaled from its value for
$J_1=1$ by a factor of $0.4$ with surprising accuracy. The scaling of the
spectrum as a whole is not as good, especially the three levels $(\pi/2,0)$,
$(\pi,0)$ and $(\pi,\pi/2)$ are shifted with respect to the bottom of the
band. But the order of the levels coincides nearly and the effects which have
been discussed above may also be observed in Fig.\ 2c. Especially, the
appearance of a flat band region, i.e. the decrease of the energy difference  
$\varepsilon(\pi,0)-\varepsilon(\pi/2,0)$ is visible. So, we can conclude that
the results for $J_1=1$ are already responsible for the realistic value
$J_1=0.4$ if we scale the energies by a factor of roughly $0.4$. Such a
scaling is already known for the case without additional hopping parameters
($t_2=t_3=0$) [11-15, 20] where the scaling is fulfilled with a
slightly higher accuracy than may be observed here. Furthermore, comparing
Figs.\ 2b and 2c with former results \cite{2,z11} we see that the additional
hopping parameters do not increase the bandwidth. That is in contradiction to
the work \cite{z13} where it was conjectured that the inclusion of the $t_2$-
and $t_3$-terms leads to an increase of the bandwidth up to a factor of four. 
\par
Finally, in Fig.\ 3 we represent the results for finite temperature $T= 1$
and zero frustration. As can be expected, in some sense the 
inclusion of temperature simulates the effect of doping by 
decreasing the antiferromagnetic correlation length in a similar way as does
the frustration. But now, the exact diagonalization method is not
applicable. The dispersion in Fig.\ 3 is similar to that one of Fig.\ 1b. In
particular, the extended saddle point between $(\pi/2,0)$ and $(\pi,0)$ can be
observed. There is one important difference that the saddle point in Fig.\ 3
lies much higher with respect to the band bottom than in Fig.\ 1b. That may
have consequences for the position of the Fermi level if we fill the band with
holes. That question, however, is beyond the scope of the present
investigation.
\section{Conclusion}
The ARPES experiments show different dispersions in the insulating and
optimally doped cuprates. One observes an isotropic band minimum but no flat
band in the insulating compound and a flat band region in the optimally doped
ones. We have shown here that the isotropic band minimum of the undoped case
may be explained due to additional hopping parameters in a $t$-$J$-like
Hamiltonian which naturally appear
in a proper reduction scheme from the three band Hubbard model. The doping
changes the spin background and in the present study we have assumed that it
may be simulated by a frustration term in the exchange energy. We have seen
that frustration (doping) leads to a non-rigid change of the bands such that a
flat band region appears.
\par
Let us emphasize  the importance of taking into account the
$t_2$- and $t_3$-hopping-terms. As mentioned above, these hoppings lead to the 
isotropization of the band bottom in the 
non-frustrated case. But these terms are also 
important for an adequate description of the spectrum in the vicinity of
the  
point $(\pi,0)$. As it was shown in \cite{z11} the frustration  leads  to 
a flat band 
region with an extended saddle point near the point $(\pi,0)$ also in the case
of the 
$t$-$J_1$-$J_2$ model. But then, in contrast to Fig.\ 1b, this saddle point is
extended in the direction $(\pi,0)$-$(\pi,\pi/2)$ and not in the direction
$(\pi/2,0)$-$(\pi,0)$ as in the experiment. So, only the 
simultaneous consideration of $t_2$-, $t_3$-hoppings and of  
frustration leads to the extended saddle point in the direction
$(\pi/2,0)$-$(\pi,0)$. 
\par
In the present study we proposed to explain the different dispersion in
insulating and doped cuprates by one characteristic set of hopping-parameters
and a non-rigid band behaviour. Since up to now the experiments in the
insulating and doped cases have been done at different compounds, another
theoretical approach is possible as well: different compounds lead to
different hopping parameters and to different one-hole dispersions which behave
more or less rigid with respect to doping. Such a point of view can be found
in \cite{z10} and \cite{Apex}. In the latter work it was proposed that the
microscopic hopping parameters are influenced especially by the position of
the apex-oxygen. Which of the two approaches is the correct one can be decided
in the end only by an experiment which shows explicitely the difference in the
dispersion of the undoped and doped case of one substance. \\

\vspace*{.1cm}

{\Large \bf ACKNOWLEDGEMENT}\\

\vspace*{.1cm}

The authors thank Johannes Richter for a lot of stimulating discussions and
for continued interest in the work. 
We thank V.M.\ Beresovsky that we could use his code to calculate the
spin-spin correlation functions. 
One of the authors (A.F.B) thanks Prof.\ Esch\-rig and the Max-Planck
Arbeitsgruppe in Dresden for hospitality where part of the work has been
carried out. Furthermore, we acknowledge the financial support of the INTAS
organization under project-number INTAS-93-285 and of the Deutsche
Forschungsgemeinschaft project-number Ri 615/1-2.
\\

\newpage
{\LARGE \bf APPENDIX}\\

\vspace*{.1cm}

The procedure to calculate the matrix elements $S^{a b} (k)$ (\ref{7}) and
$H^{a b}(k)$ (\ref{8}) is close to that one given in \cite{z11}, and we
present only the final expression. Let us introduce the following notations
for static correlation functions of Hubbard operators with noncoinciding
indices: 
\begin{eqnarray}
Z_{a} &=& \sum_s \left< X_i^{\sigma s} X_{i+a}^{s \sigma} \right >
\; , \nonumber \\
D_{a,b} &=& \sum_{s_1 s_2} \left< X_i^{\sigma s_1} X_{i+a}^{s_1 s_2}
X_{i+b}^{s_2 \sigma} \right >
\; , \nonumber \\
V_{a,b,c} &=& \sum_{s_1 s_2 s_3} \left< X_i^{\sigma s_1} X_{i+a}^{s_1 s_2}
X_{i+b}^{s_2 s_3} X_{i+c}^{s_3 \sigma} \right > \; .
\nonumber
\end{eqnarray}
These expressions can be further reduced to static spin-spin correlation
functions (for details see \cite{z11}). A straightforward calculation gives for
the overlap matrix:
\begin{equation}
\begin{array}{lll}
S^{aa}= 1/2  &(a=0,\dots,4)&  \\
S^{0a}= Z_a  &(a=1,\dots,4)&  \\
S^{ab}= D_{a,a-b} &(a,b=1,\dots,4) & (a\neq b) \; ,
\end{array}
\nonumber
\end{equation}
and we see that it doesn't depend explicitely on the momentum $k$. 
Analogously, the Hamilton matrix may be expressed as:
\begin{eqnarray}
\underline{H}(k) &=& \underline{K} (k) + \underline{E}  \nonumber \\
K^{00}(k)&=&\sum_{l} t_{l} \ Z_{l} \ \mbox{e}^{ikl} \; ,
\nonumber \\
K^{0a}(k)&=&\frac{1}{2} t_{a} \ \mbox{e}^{ika}
+ \sum_{l}\bar{\delta}_{al} \ t_{l} \ D_{l,l-a} \ \mbox{e}^{ikl}
\; ,\nonumber \\
K^{aa}(k)&=& t_{a} \ Z_{a} \ (\mbox{e}^{ika}+\mbox{e}^{-ika})
+\sum_{l}\bar{\delta}_{al} \ \bar{\delta}_{-a,l} \ t_{l} \ V_{a,a+l,l}
\ \mbox{e}^{ikl} \; ,
\nonumber \\
K^{ab}(k)&=&\bar{\delta}_{-a,b} \ t_{b} \ Z_{a} \ \mbox{e}^{ikb}
+ t_{a} \ Z_{b} \ \mbox{e}^{-ika} + t_{b-a} \ Z_{b-a} \ \mbox{e}^{ik(b-a)}
\nonumber \\
&+ & \sum_{l}
\bar{\delta}_{-a,l} \
\bar{\delta}_{bl} \
\bar{\delta}_{b-a,l} \
t_{l} \ V_{a,a+l,a+l-b} \
\mbox{e}^{ikl} \; ,
\nonumber \\
\underline{E}&=&(J_1+J_2) \ \underline{S} + \underline{\tilde{E}} \; ,
\nonumber \\
\tilde{E}^{00}&=& -\frac{1}{2}\sum_{m} J_{m} \ Z_{m} \; ,
\nonumber \\
\tilde{E}^{0a}&=& -\sum_{m} J_{m}
\left( \frac{\delta_{am}}{4}+\frac{\bar{\delta}_{am}}{2} D_{m,m-a}
\right) \; ,
\nonumber \\
\tilde{E}^{aa}&=& -\sum_{m} J_{m}
\left( \frac{\delta_{am}}{2} \ Z_{m} \
+ \bar{\delta}_{am} \left( Z_{m}-\frac{Z_{m-a}}{2} \right)
\right) \; ,
\nonumber \\
\tilde{E}^{ab}&=& \frac{1}{2}\sum_{m} J_{m} \left[
-\delta_{bm} \ Z_{a-b} -\delta_{am} \ Z_{b}
+\delta_{m-a,-b} (Z_{a-b} - Z_{a}) \right. \nonumber \\
&-&
\left. \bar{\delta}_{am} \ \bar{\delta}_{bm}
\ V_{a-m,a,a-b}
+ \bar{\delta}_{am} \ \bar{\delta}_{m-a,-b}
(V_{m,a,a-b} - V_{-m,a-m,a-m-b}) \right]
\; ,
\nonumber 
\end{eqnarray}
where $\delta_{am}=1$ for $a=m$ and zero else, and
$\bar{\delta}_{am}=1-\delta_{am}$. The indices $a$ and $b$ ($a \neq b$) run
over $1$ to $4$ and the 
above summation on $l$ and $m$ are restricted to that neighbours which are
defined in the kinetic (\ref{1a}) and exchange (\ref{1b}) parts of the
Hamiltonian, respectively.

\newpage
\vspace*{4cm}
\begin{center}
{\large\bf FIGURE CAPTIONS}
\end{center}
\vspace*{2cm}
\noindent
\begin{tabular}{ll}
FIG.\ 1:  &
\parbox[t]{12cm}
{Quasiparticle dispersion along the line $M$-$G$-$X$-$M$ and contour plot for
$J_1=t_1=1, t_2=-0.08, t_3=0.15, T=0.2$ without frustration (a) and for
$J_2=0.4$ (b). The points in $k$-space mean: $M=(\pi,\pi), G=(0,0)$ and 
$X=(\pi,0)$. The circles are the result of the exact diagonalization of a
$4*4$ lattice at $T=0$ with total spin
$S_{tot}=1/2$. } \\
\end{tabular}

\vspace*{2cm}
\noindent
\begin{tabular}{ll}

FIG.\ 2:  &
\parbox[t]{12cm}
{Energy difference between the lowest state of one hole with total spin
$S_{tot}=1/2$ 
for several momenta and the groundstate without any hole in
dependence on the frustration within
the Green's function method (GF) for $J_1=1$ (a), and for the $4*4$ lattice
for $J_1=1$ (b) and $J_1=0.4$ (c). The hopping parameters are the same as in
Fig.\ 1. }\\
\end{tabular}

\vspace*{2cm}
\noindent
\begin{tabular}{ll}
FIG.\ 3: &
\parbox[t]{12cm}
{Dispersion and contour plot for $J_2/J_1=0$ and finite temperature $T=1$. The
other parameters are as in Fig.\ 1.} \\
\end{tabular}

\end{document}